\documentclass[a4paper]{amsart}

\usepackage{amsmath}
\usepackage{amssymb}
\usepackage{amsthm}
\usepackage{amsfonts}
\usepackage{amstext}
\usepackage{amsopn}
\usepackage{amsxtra}
\usepackage{mathrsfs}
\usepackage{color}      
\usepackage{verbatim}   
\usepackage{dsfont}

\definecolor{dgreen}{rgb}{0,0.8,0.2}

\newtheorem{lemma}{Lemma}[section]
\newtheorem{theorem}{Theorem}
\newtheorem{proposition}[lemma]{Proposition}

\newtheorem{corollary}[lemma]{Corollary}
\theoremstyle{definition}

\theoremstyle{definition}
\newtheorem{remark}[lemma]{Remark}
\theoremstyle{definition}

{\catcode `\@=11 \global\let\AddToReset=\@addtoreset}
\AddToReset{equation}{section}

\newcommand\h{{\mathsf{h}}}
\newcommand\B{{\mathscr{B}}}

\newcommand\IR{{\mathbb{R}}}

\newcommand\Ff{{\mathcal{F}}}
\newcommand\Ll{{\mathcal{L}}}
\newcommand\Nn{{\mathcal{N}}}
\newcommand\Tt{{\mathcal{T}}}
\newcommand\Xx{{\mathcal{X}}}
\newcommand\unit{\hbox{\rm 1\kern-2.8truept l}}
\newcommand\re{\Re\hbox{\rm e}}

\newcommand\ep{\varepsilon}
\newcommand\Dom{\hbox{\rm Dom}}
\newcommand{\set}[1]{\left\{\,#1\,\right\}}
\newcommand{\tr}[1]{\hbox{\rm tr}\left(#1\right)}
\newcommand\Lform{{\mathcal{L}}\kern-7.56pt\raise1.0pt\hbox{$-$}}
\newcommand{\initsp}{\mathsf{h}}

\newcommand{\be}{\begin{equation}}
\newcommand{\ee}{\end{equation}}

\begin{document}


\title[Quantum Fokker-Planck models: Lindblad and Wigner approaches]{Quantum Fokker-Planck models: the Lindblad and Wigner approaches} 

\author{A. Arnold}
\address{A. Arnold: Institute for Analysis and Scientific Computing, TU Wien,
Wiedner Hauptstr. 8, A-1040 Wien, Austria}
\email{Anton.Arnold@tuwien.ac.at}

\author{F. Fagnola}
\address{F. Fagnola: Dipartimento di Matematica, Politecnico di Milano,
Piazza Leonardo da Vinci 32, I-20133 Milano, Italy}
\email{franco.fagnola@polimi.it}

\author{L. Neumann}
\address{L. Neumann: Institute for Analysis and Scientific Computing, TU Wien,
Wiedner Hauptstr. 8, A-1040 Wien, Austria}
\email{Lukas.Neumann@tuwien.ac.at}


\begin{abstract}
In this article we try to bridge the gap between the quantum dynamical semigroup and Wigner function approaches to quantum open systems. In particular we study stationary states and  the long time asymptotics for the quantum Fokker--Planck equation. Our new results apply to open quantum systems in a harmonic confinement potential, perturbed by a (large) sub-quadratic term.
\end{abstract}


\maketitle  

\section{Quantum Fokker--Planck model} 

This paper is concerned with the mathematical analysis of quantum Fokker--Planck (QFP) models, a special type of open quantum systems that models the quantum mechanical charge-transport including diffusive effects, as
needed, {\it{e.g.}}, in the description of quantum Brownian motion \cite{Di}, quantum optics \cite{EL}, and semiconductor device simulations \cite{DNJ}. We shall consider two equivalent descriptions, the Wigner function formalism and the density matrix formalism.

In the quantum kinetic Wigner picture a quantum state is described by the real valued 
Wigner function $w(x,v,t)$, where $(x,v)\in \IR^2$ denotes the position--velocity phase space. Its time evolution in a harmonic confinement potential $V_0(x)=\omega^2\frac{x^2}{2}$ with $\omega>0$ is given by
the Wigner Fokker--Planck \cite{SaSc, ALMS, SCDM} (WFP) equation 
\begin{eqnarray}\label{AFN:WFP}
{\partial_t w}& = & \omega^2 x {\partial_v w} 
- {v \partial_x w} + Qw \,,\\
Qw &=& 2\gamma {\partial_v (v w)} + D_{pp}\Delta_v w + D_{qq}\Delta_x w 
   +2 D_{pq}{\partial_v}{\partial_x w}\,. \nonumber
\end{eqnarray}
The (real vauled) diffusion constants $D$ and the friction $\gamma$ satisfy the Lindblad condition
\be\label{AFN:Lindblad-cond}
   \Delta:=D_{pp}D_{qq}-D_{pq}^2- \gamma^2/4\ge0\,,
\ee
and $D_{pp}, D_{qq}\geq 0$. Moreover we assume that the particle mass and $\hbar$ are scaled to 1.

WFP can be
considered as a quantum mechanical generalization of the usual kinetic Fokker--Planck equation (or Kramer's equation),
to which it is known to converge in the classical limit $\hbar \to 0$, after an appropriate rescaling of the
appearing physical parameters \cite{Bo, LiPa93}. The WFP equation has been partly derived in Ref. \cite{CEFM}
as a rigorous scaling limit for a system of particles interacting with a heat bath of phonons.

In recent years, mathematical studies of WFP type equations mainly focused on the Cauchy problem 
(with or without self-consistent Poisson--coupling) \cite{ADM1, ADM2, ALMS, ArSp, CLN, Ar2008}. 
In the present work we shall be concerned with the steady state problem for the WFP equation and the large-time convergence to such steady states. 
Stationary equations for quantum systems, based on the Wigner formalism, seem to be rather difficult. 
For a purely quadratic confinement potential, this problem was dealt with in Ref. \cite{SCDM} using PDE--tools. The extension to harmonic potentials with a small, smooth perturbation was recently obtained in Ref. \cite{AGGS} using fixed point arguments and spectral theory. Here we consider large perturbations of the harmonic potential. To this end we shall work in the density matrix formalism, using tools from operator theory.
\medskip

In the density matrix formalism a quantum state is described by a density matrix $\rho\in\mathscr T^+_1(\mathsf{h})$, the cone of positive trace class operators on some Hilbert space $\mathsf{h}$. 
Its time evolution is governed by the linear QFP equation or master equation
\be\label{AFN:master-eq}
\frac{d\rho_t}{dt}={\Ll}_*(\rho_t)\,,
\ee
with the Lindbladian
\begin{eqnarray}\label{AFN:Lindbladian}
{\Ll}_*(\rho) 
& = & -\frac{i}{2} \left[ p^2+\omega^2 q^2 +V(q), \rho\right]  
- i \gamma\left[ q, \{p,\rho\}\right]\notag \\
& - & D_{qq}[p,[p,\rho]] - D_{pp}[q,[q,\rho]] 
+2D_{pq}[q,[p,\rho]]\,,
\end{eqnarray}
where $V(q)$ is the perturbation of the harmonic potential.

Global in time solutions to such master equations were
established in Ref. \cite{ArSp} (nonlinear QFP--Poisson equation) starting 
from the construction of the associated minimal quantum dynamical semigroup 
(QDS) \cite{Da}. 

General methods for the study of quantum master equations 
and their large time behavior, including the existence of steady states and 
convergence towards them were developed in Quantum Probability.

Applicable sufficient conditions for proving 
uniqueness, i.e. trace preservation, of the solution 
obtained by the minimal semigroup method were given in  
Ref. \cite{ChFa} (see also Ref. \cite{GQ}). A criterion based on a non-commutative 
generalization of Liapounov functions for proving the existence of steady 
states was developed in Ref. \cite{FaRe01}. The support of steady states 
and decomposition of a quantum Markov semigroup into its transient and 
recurrent components were studied in Ref.~\cite{FaRe02} and \cite{VU}. 
When the support of a steady state is full, i.e. it is faithful, uniqueness 
of steady states and convergence towards steady states can be deduced 
from simple algebraic conditions based on commutators of 
operators appearing in a Lindblad form representation of the master 
equation (see \cite{Fr} for bounded and \cite{FaRe98}
for unbounded operators).
Many of these methods generalize those of stochastic analysis in the 
study of classical Markov semigroups and processes. We refer to the 
lecture note \cite{FaRe06} for a comprehensive account.

In this paper we study the master equation 
(\ref{AFN:master-eq}) by the above methods. We first prove the 
existence and trace preservation (i.e. uniqueness) of solutions 
and then the existence of a steady state. If the diffusion 
constants $D_{pp}, D_{pq}, D_{qq}$, and the friction $\gamma$ 
satisfy the Lindblad condition (\ref{AFN:Lindblad-cond}) with 
the {\em strict} inequality, we prove that this quantum Markovian 
evolution is irreducible in the sense of Ref. \cite{FaRe03}. 
As a consequence, steady states must be faithful and one can 
apply simple commutator conditions on the operators in the 
GKSL representation to establish uniqueness of the steady 
state and large time convergence towards this state. 

When $\Delta=0$
we conjecture that (see Sect.~\ref{AFN:sect-limiting}), unless $V$ 
is zero and the limiting conditions $D_{qp}=-\gamma D_{qq}$, 
$D_{pp}=\omega^2 D_{qq}$ are satisfied, the quantum Markov 
semigroup is still irreducible. But the invariant subspace 
problem that has to be solved for proving this becomes very 
difficult and we were not been able to solve it. 

The paper is organized as follows: In Section~\ref{AFN:section2} we review the equivalence of the kinetic Wigner formalism and the Lindblad approach to open quantum systems. Some technical preliminaries are presented in \S\ref{AFN:sect-alg-comp} and \ref{domain}. In \S\ref{QDS} we construct the minimal QDS for \eqref{AFN:master-eq}, \eqref{AFN:Lindbladian} with external potentials that grow at most subquadratically. The markovianity of the semigroup is proved in \S\ref{s.6}. This yields uniqueness and mass--conservation of the solution to \eqref{AFN:master-eq}, \eqref{AFN:Lindbladian}. In \S\ref{steady} we establish the existence of a steady state and in \S\ref{AFN:sect-irreduc} we prove that the solution converges to this unique steady state for arbitrary initial data provided the Lindblad condition~\eqref{AFN:Lindblad-cond} is fulfilled with strict inequality. The limiting case $\Delta=0$ is studied in \S\ref{AFN:sect-limiting}.



\section{Passage from the Wigner equation to the master equation} \label{AFN:section2}

In this section we show how to pass from the Wigner language to the GKLS (Gorini, Kossakowski, Sudarshan \cite{GKS}; Lindblad \cite{Lind}) language.
In order to keep the notation simple, we shall confine our presentation to the one dimensional case. However, the results extend to higher dimensions. 
The underlying Hilbert space of our considerations is ${\mathsf{h}}=L^2(\IR)$. We denote by $q$ and $p$ the standard position and momentum operators ($p=-i\partial_x$). They satisfy the canonical commutation relation (CCR) $[q,p]=i\unit$.

The Wigner function $w(x,v,t)$ of a state $\rho_t={\mathcal T}_{*t}(\rho)$, is (up to normalization) the anti Fourier trans\-form of 
\be\label{AFN:transfo}
\varphi(\xi,\eta)=\hbox{\rm tr}
\left(\rho_t\,\hbox{\rm e}^{-i(\xi q+\eta p)}\right)\,.
\ee
Using \eqref{AFN:transfo} we shall now transform the WFP equation \eqref{AFN:WFP} into an evolution equation for the corresponding density matrix $\rho$.\\
As a consequence of the CCR we have
\begin{eqnarray}
\hbox{\rm e}^{-i(\xi q+\eta p)} 
& = & \hbox{\rm e}^{-i\xi q}\hbox{\rm e}^{-i\eta p}\hbox{\rm e}^{{i\xi\eta}/{2}} \label{AFN:eq-BCH-} \,,\\
\hbox{\rm e}^{-i(\xi q+\eta p)} 
& = &\hbox{\rm e}^{-i\eta p}\hbox{\rm e}^{-i\xi q}\hbox{\rm e}^{-{i\xi\eta}/{2}} \label{AFN:eq-BCH+}\,,\\
\hbox{\rm e}^{-i(\xi q+\eta p)} 
& = &\hbox{\rm e}^{-i\eta p/2}\hbox{\rm e}^{-i\xi q}\hbox{\rm e}^{-i\eta p/2}\,.\label{AFN:eq-BCH=} 
\end{eqnarray}
Assuming that $\rho$ is sufficiently regular, by differentiating (\ref{AFN:eq-BCH-}) 
and (\ref{AFN:eq-BCH+}) and using the cyclic property of the trace, we find
\begin{eqnarray*}
\partial_\xi\varphi(\xi,\eta) 
&=& -i\,\tr{\rho\, q\, \hbox{\rm e}^{-i(\xi q+\eta p) }} +\frac{i\eta}{2}\varphi(\xi,\eta)\,,\\
\partial_\xi\varphi(\xi,\eta) 
&=& -i\,\tr{q\,\rho\,  \hbox{\rm e}^{-i(\xi q+\eta p) }} -\frac{i\eta}{2}\varphi(\xi,\eta)\,.
\end{eqnarray*}
Subtracting, and respectively, summing the above equations we have 
\begin{eqnarray*}
\eta\varphi(\xi,\eta) & = & -\tr{[q,\rho]\,\hbox{\rm e}^{-i(\xi q+\eta p)}} \label{AFN:eq-eta-phi}\,, \\
\partial_\xi\varphi(\xi,\eta) & = & -\frac{i}{2}\tr{\{q,\rho\}\,\hbox{\rm e}^{-i(\xi q+\eta p)}}\,.
\end{eqnarray*}
In a similar way we obtain formulae for the products and derivatives with respect to $\xi$:
\begin{eqnarray*}
\xi\varphi(\xi,\eta)  & = &  \tr{[p,\rho]\hbox{\rm e}^{-i(\xi q+\eta p)}} \label{AFN:eq-xi-phi}\,, \\
\partial_\eta\varphi(\xi,\eta) & = & -\frac{i}{2}\tr{\{p,\rho\}\,\hbox{\rm e}^{-i(\xi q+\eta p)}}\,. 
\end{eqnarray*}

\medskip
The Wigner function is the anti Fourier transform of $\varphi$
\[
w(x,v) = \left(\frac{1}{2\pi}\right)^2\int_{\IR^2}\hbox{\rm e}^{i(\xi x+\eta v)}\varphi(\xi,\eta)d\xi d\eta\,.
\]
The factor is chosen such that the total mass is given by
$$
m=\tr{\rho}=\varphi(0,0)=\int_{\IR^2}w(x,v)dxdv\,.
$$
Integrating by parts (and again assuming sufficient regularity and decay) we obtain
\begin{eqnarray*}
x w(x,v) & = & \frac{1}{2\pi}\int_{\IR^2}x\hbox{\rm e}^{i(\xi x+\eta v)}\varphi(\xi,\eta)d\xi d\eta \\
& = & \frac{1}{2\pi}\int_{\IR^2}-i\left(\partial_\xi\hbox{\rm e}^{i(\xi x+\eta v)}\right)\varphi(\xi,\eta)d\xi d\eta  \\
& = & \left[\frac{-i\hbox{\rm e}^{i(\xi x+\eta v)}\varphi(\xi,\eta)}{2\pi}\right]_{-\infty}^{+\infty}
+\frac{i}{2\pi}\int_{\IR^2}\hbox{\rm e}^{i(\xi x+\eta v)}\left(\partial_\xi\varphi(\xi,\eta)\right)d\xi d\eta \\
& = & \frac{1}{2\pi}\int_{\IR^2}\hbox{\rm e}^{i(\xi x+\eta v)}
\left(\frac{1}{2}\tr{\{q,\rho\}\,\hbox{\rm e}^{-i(\xi q+\eta p)}}\right)d\xi d\eta\,.
\end{eqnarray*}
In a similar way one can calculate
\begin{eqnarray*}
vw(x,v) & = & \frac{1}{2\pi}\int_{\IR^2}\hbox{\rm e}^{i(\xi x+\eta v)}
\left(\frac{1}{2}\tr{\{p,\rho\}\,\hbox{\rm e}^{-i(\xi q+\eta p)}}\right)d\xi d\eta \,,\\
\partial_x w(x,v) & = & \frac{1}{2\pi}\int_{\IR^2}\hbox{\rm e}^{i(\xi x+\eta v)}
\left(\tr{i[p,\rho]\,\hbox{\rm e}^{-i(\xi q+\eta p)}}\right)d\xi d\eta \,,\\
\partial_v w(x,v) & = & \frac{1}{2\pi}\int_{\IR^2}\hbox{\rm e}^{i(\xi x+\eta v)}
\left(\tr{-i[q,\rho]\,\hbox{\rm e}^{-i(\xi q+\eta p)}}\right)d\xi d\eta \,.
\end{eqnarray*}
The above formulae lead to the following dictionary for translating a master equation 
from the Wigner function language to the GKSL language:
\[
\begin{array}{|c|c|}
\hline 
\hbox{\rm transformation on $w$} & \hbox{\rm transformation on $\rho$}  \\ \hline
x w & \frac{1}{2}\{q,\rho\} \\
v w & \frac{1}{2}\{p,\rho\} \\
\partial_x w & i[p,\rho] \\
\partial_v w & -i[q,\rho] \\
\hline
\end{array}
\]
For the terms appearing in the WFP equation we have:
\[
\begin{array}{|c|c|}
\hline 
\hbox{\rm transformation on $w$} & \hbox{\rm transformation on $\rho$}  \\ \hline
x \partial_v w & -\frac{i}{2}\{q,[q,\rho]\} = -\frac{i}{2}[q^2,\rho]\\
v \partial_x w & \frac{i}{2}\{p,[p,\rho]\} = \frac{i}{2}[p^2,\rho]\\
\partial_v (vw) & -\frac{i}{2}[q,\{p,\rho\}] = -\frac{i}{2}\{p,[q,\rho]\} +\rho \\
\Delta_v w & -[q,[q,\rho]]\\
\Delta_x w & -[p,[p,\rho]]\\
\partial_x\partial_v w &  [p,[q,\rho]] \\
\hline
\end{array}\ 
\]

\bigskip

Using this dictionary we find the following GKSL form ($\frac{d\rho_t}{dt}={\Ll}_*(\rho_t)$) 
of the linear QFP equation
\begin{eqnarray}\label{e:11a}
{\Ll}_*(\rho) 
& = & -\frac{i}{2} \left[ p^2+\omega^2 q^2 , \rho\right]  
- i \gamma\left[ q, \{p,\rho\}\right] \label{AFN:eq-QFP}\\
& - & D_{qq}[p,[p,\rho]] - D_{pp}[q,[q,\rho]] 
+2D_{pq}[q,[p,\rho]]\,.\nonumber
\end{eqnarray}
This corresponds to choosing $\lambda=\mu=\gamma$ in (3.8) of Ref. \cite{SaSc} (see also \cite{ISSSS}).

The dual equation of (\ref{e:11a}) with an added perturbation potential $V$ reads
\begin{eqnarray*}
{\Ll}(A) & = & \frac{i}{2} \left[ p^2+\omega^2 q^2 + 2V(q), A\right]  
+ i \gamma\left\{ p, [q,A]\right\} \label{AFN:eq-QFP-GKSL}\\
& - & D_{qq}[p,[p,A]] - D_{pp}[q,[q,A]] +2D_{pq}[q,[p,A]]\,,
\quad A\in\B(\h)\,. 
\end{eqnarray*} 
It can be written \cite{ALMS} in (generalised) GKSL form like
\begin{equation}\label{e:12a}
\Ll(A) = i[H,A] -\frac{1}{2}\sum_{\ell=1}^2
\left( {L^*_\ell L_\ell A} - 2 L^*_\ell A L_\ell + A L^*_\ell L_\ell\right)
\end{equation}
with the ``adjusted'' Hamiltonian
\[
H= \frac{1}{2}\left(p^2+\omega^2 q^2 + \gamma(pq+qp)\right) + V(q)\,, 
\]
and the Lindblad operators $L_1$ and $L_2$ given by
\begin{equation} \label{AFN:Lindblad-Operators}
L_1= \frac{-2D_{pq}+i\gamma}{\sqrt{2 D_{pp}}} p + \sqrt{2D_{pp}}\,q\,, \qquad
L_2= \frac{2\sqrt{\Delta}}{\sqrt{2D_{pp}}} \,p\,.
\end{equation}



\section{Key inequalities for the existence of a steady state}\label{AFN:sect-alg-comp}

We aim at applying the criterion for existence of a normal 
invariant state \cite{FaRe01} by Fagnola and Rebolledo. To this end we have to find a positive operator $X$ and an operator $Y$ bounded from below, satisfying
\[
{\mathcal{L}}(X)\le - Y\,,
\]
which in addition both have finite dimensional 
spectral projections associated with intervals $]-\infty, \Lambda]$.
To illustrate the technique we first present the computation for the harmonic potential only. The perturbation potential $V(q)$ will be added later on.

Consider the Lindbladian 
\begin{eqnarray*}
{\Ll}_*(\rho) 
& = & - \frac{i}{2} \left[ p^2+\omega^2 q^2 , \rho\right]  
- i \gamma\left[ q, \{p,\rho\}\right] \\
& - & D_{qq}[p,[p,\rho]] - D_{pp}[q,[q,\rho]] + 2D_{pq}[q,[p,\rho]]
\end{eqnarray*}
with dual\cite{ISSSS}
\begin{eqnarray*}
{\Ll}(X) 
& = & \frac{i}{2} \left[ p^2+\omega^2 q^2 , X\right]  
+ i \gamma\left\{ p, [q,X]\right\} \\
& - & D_{qq}[p,[p,X]] - D_{pp}[q,[q,X]] 
+2D_{pq}[q,[p,X]]\,.
\end{eqnarray*}

Straightforward computations with the CCR $[q,p]=i\unit$ yield

\begin{lemma} The following formulae hold for $f$, $g$ smooth:
\begin{eqnarray*}
{\mathcal{L}}(f(p)) & = & -\frac{\omega^2}{2}\left(q f^\prime(p)+f^\prime(p)q\right)
-2\gamma p f^\prime(p) + D_{pp}f''(p) \,,\\
{\mathcal{L}}(g(q)) & = & \frac{1}{2}\left(p g^\prime(q)+g^\prime(q)p\right)
+ D_{qq} g''(q) \,, \\
{\mathcal{L}}(pq+qp) & = & 2\left(p^2 -\omega^2 q^2\right) 
-2\gamma\left(pq+qp\right) + 4 D_{pq}\,.
\end{eqnarray*}
\end{lemma}

This suggests looking for $X,\,Y$ given by second order polynomials in $p$ and $q$ ({\it{i.e.}} $f(p)=p^2,\,g(q)=q^2$). Therefore we start studying some algebraic properties of these operators:

\begin{lemma}\label{AFN:lem-op-rs-positive}
For all $r,s>0$ such that $rs>1$ the operators 
\[
rp^2 - (pq+qp) + sq^2, \qquad rp^2 + (pq+qp) + sq^2
\]
are strictly positive and have discrete spectrum. Moreover all spectral 
projections associated with bounded intervals are finite dimensional.
\end{lemma}

\begin{proof}
Let $r_0,s_0>0$ be such that $r_0<r$, $s_0<s$ and $r_0 s_0=1$. 
Then
\[
\left |\sqrt{r_0}\, p-\sqrt{s_0}\, q\right|^2 = r_0 p^2 - (pq+qp) + s_0 q^2\ge 0\,.
\]
It follows that 
\[
rp^2 - (pq+qp) + sq^2 = \left |\sqrt{r_0}\, p-\sqrt{s_0}\,q\right|^2 + (r-r_0) p^2 + (s-s_0) q^2\,.
\]
Therefore the resolvent of $rp^2 - (pq+qp) + sq^2$ is dominated by the resolvent of 
a multiple (indeed $\min\left\{\,(r-r_0),\,(s-s_0)\,\right\}$) of the 
number operator $\frac12({p^2+q^2}-1)$. Since the latter is compact, also the resolvent of $rp^2 - (pq+qp) + sq^2$ is compact and self-adjoint. Hence it has a discrete spectrum that might only accumulate at $0$.

The proof for the second operator is the same. 
\end{proof}

We choose $X$ of the form 
\begin{equation}\label{AFN:eq-op-X}
X=r p^2 + (pq+qp) + s q^2
\end{equation}
and compute
\begin{eqnarray*}
& & {\mathcal{L}}(r p^2 + (pq+qp) + s q^2) = -2(2\gamma r -1)p^2 -2\omega^2 q^2  \\
& & + (s-2\gamma-\omega^2 r)(pq+qp) +2r D_{pp} +4 D_{pq} +2s D_{qq}\, .
\end{eqnarray*}
The required conditions on $X$ and $Y$ ({\it{i.e.}} $X>0,\;\mathcal{L}(X)\le -Y$) hold if
\begin{eqnarray*}
 rs & > &  1 \,,\\
 4\omega^2(2\gamma r -1) & > & \left| s-2\gamma-\omega^2 r \right|^2 \, .
\end{eqnarray*}
Letting $r$ and $s$ go to infinity with $s-2\gamma-\omega^2 r$ constant (that 
can be $0$, for simplicity), it is clear that, {when $\gamma>0$}, 
we can find $r$ and $s$ large enough satisfying the above condition.
We take, {\it e.g.} any $r>(2\gamma)^{-1}$ and $s=2\gamma+\omega^2 r$ since
\[
rs = 2\gamma r + \omega^2 r^2 > 1 + \frac{\omega^2}{4\gamma^2} > 1\,.
\]
$Y$ will be chosen later in Theorem~\ref{AFN:th-exists-invariant state}.

\bigskip 

We now add the perturbation potential.
Let $V:\mathbb{R}\to\mathbb{R}$ be a smooth function satisfying a growth condition
like 
\be\label{AFN:potential-cond}
\left | V^\prime (x) \right| \le g_V \left(1+|x|^{2}\right)^{\alpha/2}
\ee
with $g_V>0$ and $0\le\alpha<1$.
Hence, this perturbation potential is strictly sub-quadratic. It gives rise to one additional term in $\mathcal{L}(X)$, namely:
\begin{eqnarray*}
i\left[\, V(q), p^2\,\right] 
& = & -\left(p V^\prime(q)+V^\prime(q)p\right) \,,\\
i\left[\, V(q), pq+qp\,\right] & = & -2q V^\prime(q)\,.
\end{eqnarray*}
Therefore we find now
\begin{eqnarray}\label{e:16}
& & {\mathcal{L}}(r p^2 + (pq+qp) + s q^2) \\
& & = -2(2\gamma r -1)p^2 -2\omega^2 q^2 + (s-2\gamma-\omega^2 r)(pq+qp)  \nonumber\\
& & \quad -r\left(p V^\prime(q)+V^\prime(q)p\right) - 2q V^\prime(q) 
    +2r D_{pp} +4 D_{pq} +2s D_{qq} \,.\nonumber
\end{eqnarray}
Note that (due to the positivity of $|\epsilon^{1/2}p \pm \epsilon^{-1/2}V^\prime(q)|^2$) 
\[
-\left(  \epsilon p^2 + \frac{1}{\epsilon} (V^\prime(q))^2\right) \le 
p V^\prime(q)+V^\prime(q)p \le \epsilon p^2 + \frac{1}{\epsilon} (V^\prime(q))^2\,.
\]
Therefore, playing on the $\ep$ and the bound on the derivative of $V$, 
we can find the needed inequality ${\mathcal{L}}(X)\le - Y$.
This will be used in \S \ref{steady} to prove the existence of a steady state.



\section{Domain problems}\label{domain}

First we define the number operator $N:=\frac12(p^2+q^2-1)$ on $\h$ with
$$
  \Dom(N) = \set{u\in\initsp\,\Big|\, Nu\in\h}
  = \set{u\in\initsp\,\Big|\, p^2u,\,q^2u\in\h}\,,
$$
where the last equality follows easily from $\|Nu\|_\h^2<\infty$ by an integration by parts. $C_c^\infty(\IR)$ is a core for $N$ (cf. Ref. \cite{ReSi86}, {\it{e.g.}}).
Let $X$ be the self-adjoint extension of \eqref{AFN:eq-op-X} (still denoted by $X$). $\Dom(X)$ is its maximum domain and $C_c^\infty(\IR)$ is a core for $X$.

The position and momentum operators are defined on $\Dom(N^{1/2})$.
Both $q$ and $p$ have, by Nelson's analytic vector 
theorem, self-adjoint extensions that will be still denoted by $q$ and $p$.\\

First we shall compare the domains of $N$ and $X$. To this end we need

\begin{lemma}\label{AFN:lem-X-R-S}
Let $r,\,s>0$ with $rs>1$ and define
\[
R := r^{1/2}p+r^{-1/2}q,\qquad S:=s^{1/2}q+s^{-1/2}p\,.
\]
Then, for all $u\in C_c^\infty(\IR)$ the following identities hold
\begin{eqnarray*}
\langle u, X^2 u \rangle = \left(s-\frac{1}{r}\right){^2}\left\langle u, q^4 u\right\rangle
  & + & 2\left(s-\frac{1}{r}\right)\langle u, qR^2q u \rangle\\
  & + & \langle u, R^4 u \rangle -2(rs-1)\Vert u\Vert^2\,,  \\
  &&\\
\langle u, X^2 u \rangle = 
  \left(r-\frac{1}{s}\right){^2}\left\langle  u, p^4 u\right\rangle
  & + & 2\left(r-\frac{1}{s}\right)\langle u, pS^2p u \rangle\\
  & + & \langle u, S^4 u \rangle -2(rs-1)\Vert u\Vert^2\,.
\end{eqnarray*}
\end{lemma}

\begin{proof} Since $u$ belongs to the domain of any monomial in $p$ and $q$ 
the proof can be reduced to the algebraic computation avoiding writing $u$'s.

Starting from the identity $X=\theta q^2+R^2$ with $\theta=s-1/r$ we have
\[
X^2 = \theta^2 q^4 + \theta \left(q^2R^2 + R^2 q^2\right)  + R^4\,.
\]
The mixed product term can be written in the form
\begin{eqnarray*}
q^2R^2 + R^2 q^2 & = & q R^2 q + q\left[q,R^2\right] + qR^2q+ \left[R^2,q\right]q \\
   & = & 2 q R^2 q + qR\left[q,R\right]+ q\left[q,R\right]R + \left[R,q\right]Rq + R\left[R,q\right]q \\
   & = & 2 q R^2 q + 2ir^{1/2}qR - 2ir^{1/2}Rq \\
   & = & 2 q R^2 q  +2ir^{1/2} \left[q,R\right] \\
   & = & 2 q R^2 q -2r\,.
\end{eqnarray*}
The conclusion is now immediate. 
\end{proof}

\medskip 
The following lemma gives similar inequalities for the operators $rp^2+sq^2$ 
({\it{i.e.}} $X$ without mixed products) that will be useful in the sequel

\begin{lemma}\label{AFN:lem-r-s}
For all $r,s>0$ and $u\in C_c^\infty(\IR)$ we have
\begin{eqnarray*}
\left\langle u, \left(rp^2+sq^2\right)^2 u\right\rangle  
  & \ge & (r\wedge s)^2 \left\langle u, \left(p^2+q^2\right)^2 u\right\rangle 
  - 2\left(rs-(r\wedge s)^2\right)\Vert u\Vert^2\,,\\
\left\langle u, \left(rp^2+sq^2\right)^2 u\right\rangle 
  & \le & (r\vee s)^2\left\langle u, \left(p^2+q^2\right)^2 u\right\rangle
  + 2\left((r\vee s)^2-rs\right)\Vert u\Vert^2 \,,
\end{eqnarray*}
where $r\wedge s=\min\left\{\,r,s\,\right\}$ and $r\vee s=\max\left\{\,r,s\,\right\}$
\end{lemma}

\begin{proof} 
Indeed
\begin{eqnarray*}
\left(rp^2+sq^2\right)^2 & = & r^2p^4 +rs\left(p^2q^2+q^2p^2\right) +s^2q^4 \\
   & = & r^2p^4 + 2rs\, pq^2p+s^2q^4 -2rs \\
   & \ge & (r\wedge s)^2 \left( p^4 +2pq^2p+q^4\right) - 2rs \\
   & = & (r\wedge s)^2 \left(p^2+q^2\right)^2 - 2\left(rs-(r\wedge s)^2\right)\,.
\end{eqnarray*}
Moreover 
\begin{eqnarray*}
\left(rp^2+sq^2\right)^2 & = & r^2p^4 +rs\left(p^2q^2+q^2p^2\right) +s^2q^4 \\
   & = & r^2p^4 + 2rs\, pq^2p+s^2q^4 -2rs \\
   & \le & (r\vee s)^2 \left( p^4 +2pq^2p+q^4\right) - 2rs \\
   & = & (r\vee s)^2 \left(p^2+q^2\right)^2 + 2\left((r\vee s)^2-rs\right)\,.
\end{eqnarray*}
This completes the proof. 
\end{proof}

\begin{proposition}\label{AFN:prop-domN-domX}
The domains of the operators $N$ and $X$coincide.
\end{proposition}

\begin{proof} We first show that $\Dom(X){\subseteq}\Dom(N)$. By Lemma~\ref{AFN:lem-X-R-S}, 
for all ${u\in C_c^\infty}$ we have 
\begin{eqnarray*}
\Vert Xu \Vert^2 +2(rs-1)\Vert u\Vert^2
  & \ge & \frac{1}{2}\min\left\{\, (s-r^{-1})^2,(r-s^{-1})^2\,\right\}
         \left\langle u, (p^4+q^4)u\right\rangle \\
  & \ge & \frac{1}{4}\min\left\{\, (s-r^{-1})^2,(r-s^{-1})^2\,\right\}
         \left\langle u, (p^2+q^2)^2u\right\rangle\,,
\end{eqnarray*}
where we used the elementary inequality $(p^2+q^2)^2\le 2(p^4+q^4)$ and ${r,\,s>0;\,rs>1}$. Therefore 
we find a constant $c_1(r,s)>0$ such that
\be\label{AFN:X-estimate}
\Vert Xu \Vert^2 +2(rs-1)\Vert u\Vert^2 \ge c_1(r,s)\left\Vert (p^2+q^2)u\right\Vert^2\,.
\ee
Now, if $u\in\Dom(X)$, then there exists a sequence $(u_n)_{n\ge 1}$ is $C_c^\infty$ 
converging in norm to $u$ such that $(Xu_n)_{n\ge 1}$ converges in norm to 
$Xu$ (since $X$ is closed). The above inequality shows then that the sequence $(Nu_n)_{n\ge 1}$ 
is also norm convergent and its limit ($N$ is closed) is $Nu$.

This shows that $\Dom(X)\subseteq\Dom(N)$ and the two domains coincide since 
the opposite inclusion holds true by the construction of $X$ itself. 
\end{proof}
For future reference we briefly recall the definition of the annihilation operator $a:=\frac{1}{\sqrt{2}}(q+ip)$ and the creation operator $a^\dagger:=\frac{1}{\sqrt{2}}(q-ip)$. Using the isomorphic identification of $\mathsf{h}$ with $l^2(\mathbb{N}_0)$ (via the eigenfunctions of $N$, {\it i.e.} the Hermite functions), these operators can also be represented as 
\begin{equation*}
ae_0=0,\quad  ae_{j+1}=\sqrt{j+1}e_j,\quad a^\dagger e_{j}=\sqrt{j+1}e_{j+1}\quad\text{, for } j\in\mathbb{N}_0\, .
\end{equation*}
Here, $(e_j)_{j\geq 0}$ is the canonical orthonormal basis of $l^2(\mathbb{N}_0)$.



\section{Construction of the minimal quantum dynamical semigroup}\label{QDS}

In this section we shall establish the existence of the minimal quantum dynamical semigroup (QDS) for the Lindbladian \eqref{e:12a}. To this end we first consider the operator 
$G$, defined on $\Dom(N)$ by  
\begin{eqnarray}
G & = &-\frac{1}{2}\left(L_1^*L_1+L_2^*L_2\right) - iH =  -\left(D_{qq}+\frac{i}{2}\right)p^2 
   -\left(D_{pp}+\frac{i\omega^2}{2}\right)q^2 \notag\\
  & + & \left(D_{pq}-\frac{i\gamma}{2}\right)(pq+qp)+\frac{\gamma}{2} -iV(q)\,. \label{AFN:eq-G}
\end{eqnarray}
We suppose that the potential $V$ is twice differentiable and satisfies 
the growth condition \eqref{AFN:potential-cond}.

\begin{proposition}\label{AFN:prop-domN-domG}
The domains of the operators $N$, $G$, {and $G^*$} coincide. 
\end{proposition}

\begin{proof} 
Since $\Dom(N)\subseteq\Dom(G)$ by 
construction, it suffices to check the opposite inclusion. To this end we proceed 
as before finding an estimate of $\Vert Nu\Vert$ by the graph norm of $G$.
Putting 
\[
G_0 = -\frac{1}{2}\left(L_1^*L_1+L_2^*L_2\right) 
= {-}D_{qq}p^2 {+}D_{pq}(pq+qp) - D_{pp}q^2 +\frac{\gamma}{2}\,,
\]
we have for all $u\in C_c^\infty(\IR)$
\[
\Vert G u \Vert^2 = \Vert G_0 u \Vert^2 
      + \left\langle u, i\left[{H,G_0}\right] u\right\rangle + \Vert Hu\Vert^2\,.
\]
A straightforward computation yields
\begin{eqnarray*}
i\left[G_0,H\right] & = & -2(\gamma D_{qq}+D_{pq})p^2 + 2({\gamma D_{pp}+}\omega^2 D_{pq})q^2
    +2D_{pq}qV^\prime(q) \\
        & & {+}\left(D_{pp}{-}\omega^2 D_{qq}\right)(pq+qp) -D_{qq}\left( pV^\prime(q)+V^\prime(q)p\right)\,.
\end{eqnarray*}
The above commutator has quadratic monomials in $p$ and $q$ and other terms 
including $V^\prime(q)$ whose growth is sublinear due to our hypothesis on the 
potential. It follows then that we can find a constant $c_3>0$ such that 
\[
\left|\left\langle u, i\left[G_0,H\right] u\right\rangle \right|
\le c_3 \left\langle u, (p^2+q^2) u\right\rangle\,.
\]
Hence, for all $\epsilon>0$, by the Schwarz inequality
\[
\left|\left\langle u, i\left[G_0,H\right] u\right\rangle \right|
\le \epsilon \left\Vert (p^2+q^2) u\right\Vert^2 +{c_3^2}\epsilon^{-1}\Vert u\Vert^2\,.
\]
It follows that
\be\label{AFN:G-estimate}
\Vert G u \Vert^2 \ge \Vert G_0 u \Vert^2 
- \epsilon \left\Vert (p^2+q^2) u\right\Vert^2 -{c_3^2}\epsilon ^{-1}\Vert u\Vert^2\,.
\ee

\noindent
\underline{Case 1:}\\
Let $D_{pq}\not=0$ and $D_{pp}D_{qq}>D_{pq}^2$ (cp. to the Lindblad condition \eqref{AFN:Lindblad-cond}). Then
\[
\frac{D_{qq}}{D_{pq}}\frac{D_{pp}}{D_{pq}}=\frac{D_{pp}D_{qq}}{D_{pq}^2}
>1
\]
and
\[
G_0=-{D_{pq}}\left( \frac{D_{qq}}{D_{pq}}p^2 -(pq+qp) 
+  \frac{D_{pp}}{D_{pq}}q^2\right)+\frac{\gamma}{2}\,.
\]
Therefore $G_0$ is a multiple of an operator like those of Lemma~\ref{AFN:lem-op-rs-positive} 
(up to a constant). By inequality \eqref{AFN:X-estimate} 
we can find positive constants $c_4,\,c_5>0$ such that ${\Vert G_0 u \Vert^2 \ge c_4 
\left\Vert (p^2+q^2) u\right\Vert^2-c_5\left\Vert u\right\Vert^2}$. Therefore we derive the inequality
\be\label{AFN:G-est}
\Vert G u \Vert^2 \ge \left(c_4-\epsilon\right)\Vert (p^2+q^2) u\Vert^2-{c_6}\Vert u\Vert^2
\ee
for all $u\in C_c^\infty(\IR)$. Choosing $\epsilon=c_4/2$, we find an inequality allowing us to 
repeat the above argument for a sequence $(u_n)_{n\ge 1}$ (see the proof of Prop.~\ref{AFN:prop-domN-domX}) and prove the inclusion
$\Dom(G)\subseteq\Dom(N)$.\\

\noindent
\underline{Case 2:}\\
Let $D_{pq}=0$. The estimate \eqref{AFN:G-est} now follows from \eqref{AFN:G-estimate} and Lemma~\ref{AFN:lem-r-s}. We conclude again that $\Dom(G)\subseteq\Dom(N)$.\\

\noindent
\underline{Case 3:}\\
Let $D_{pq}\not=0$ and $D_{pp}D_{qq}=D_{pq}^2$ (and hence $\gamma=0$). To recover  \eqref{AFN:G-est} we start from 
\be\label{AFN:G-estimate-3}
\Vert G u \Vert^2 \ge \Vert H u \Vert^2 
- \epsilon \left\Vert (p^2+q^2) u\right\Vert^2 -{c_3^2}\epsilon ^{-1}\Vert u\Vert^2
\ee
(in analogy to \eqref{AFN:G-estimate}), where we have now $H=\frac12 (p^2+\omega^2q^2)+V(q)$. 
{}From Lemma~\ref{AFN:lem-r-s} we obtain 
\begin{equation*}
\begin{split}
   \|(p^2+q^2)u\|^2 & \le c_1\|\frac12(p^2+\omega^2q^2)u\|^2+c_2\|u\|^2 \\
   & \le 2c_1 (\|Hu\|^2 +\|V(q)u\|^2)+c_2\|u\|^2\,.
\end{split}
\end{equation*}
Since $V(q)$ is sub-quadratic we have
$$
  \|V(q)u\|^2 \le \epsilon \|q^2u\|^2 +{c_3^2}\epsilon ^{-1}\|u\|^2 
  \le \epsilon \|(p^2+q^2)u\|^2 +c_4\|u\|^2 \,.
$$
Hence,
\be\label{AFN:H-estimate}
   \|(p^2+q^2)u\|^2  -c_5\Vert u\Vert^2\le c_6\|Hu\|^2\,,
\ee
and we conclude by combining \eqref{AFN:G-estimate-3} and \eqref{AFN:H-estimate}.
\end{proof}
\begin{remark}\label{AFN:graph}
{}From Propositions~\ref{AFN:prop-domN-domX} and \ref{AFN:prop-domN-domG} we infer that the graph norms of {$G,\ X\ \text{and}\ N$} are equivalent. Hence $G$ is relatively bounded by $X$.
\end{remark}
The operator $G$ is clearly dissipative because 
\[
\Re \langle u, G u \rangle = -{\frac{1}{2}}\sum_{\ell=1}^2\Vert L_\ell u \Vert^2 \le 0
\]
for all $u\in\Dom(N)$. Therefore, by Prop.~3.1.15 of Ref. \cite{BrRo} it is closable and its closure is dissipative. We denote by the 
same symbol $G$ the closure. Analogously, $G^*$ is also dissipative.

Hence, the Lumer--Phillips theorem (see Theorem~3.1.16 of \cite{BrRo}, {\it{e.g.}}) yields:

\begin{proposition}\label{AFN:prop-G-generates}
The operator $G$ generates a strongly continuous contraction semigroup
$(P_t)_{t\ge 0}$ on $\initsp$.
\end{proposition}

Since $\Dom(G)=\Dom(N)\subset \Dom(L_\ell),\;\ell=1,\,2$, and since 
$$
  \langle Gv,u\rangle + \sum_{\ell=1}^2 \langle L_\ell v,L_\ell u\rangle
  +\langle v,Gu\rangle=0\quad \forall\,u,v\in \Dom(N)\,,
$$
condition {\bf (H)} of Ref. \cite{FaRe01} holds and we can construct $\Tt$, the minimal QDS \cite{Da} associated with $G$ and the $L_\ell$'s.



\section{Markovianity of the Quantum Dynamical Semigroup}
\label{s.6}

The hypotheses for constructing the minimal quantum dynamical 
semigroup with form generator 
\begin{equation}\label{AFN:eq-form-generator}
\pounds(A)[v,u] = \langle Gv,Au\rangle 
+ \sum_{\ell=1}^2\langle L_\ell v, A L_\ell u\rangle
+ \langle v,AGu\rangle , \quad A\in \B(\h)
\end{equation}
hold by Prop.~\ref{AFN:prop-G-generates}.

Now we want to show that the minimal semigroup $\Tt$ is Markov and hence mass conserving.
To this end we apply Theorem 4.4 in Ref.~\cite{ChFa}.

The algebraic computations in Sec.~\ref{AFN:sect-alg-comp} suggest to consider an  
operator $X$ of the form (\ref{AFN:eq-op-X}) with $r>(2\gamma)^{-1}$ and $s=2\gamma+\omega^2 r$ 
on the linear manifold $C_c^\infty(\IR)$. The algebraic computations of this section can be made in the quadratic form sense.

\medskip
We now check that the operator $X$ satisfies the fundamental hypothesis ${\bf C}$ of Ref. \cite{ChFa}
starting from domain properties:

\begin{proposition}\label{AFN:prop-dom-hypoC}
The following properties of $G,\,L_\ell$, and $N$ hold:
\begin{enumerate}
\item $\Dom(G)=\Dom(X)\subseteq \Dom(X^{1/2})$ and $\Dom(G)$ is a core for $X^{1/2}$\,,
\item $L_\ell(\Dom(G^2))\subseteq \Dom(X^{1/2})$ for $\ell=1,2$\,.
\end{enumerate}
\end{proposition}

\begin{proof} 
Clearly, the first part of (1) 
follows from the Propositions \ref{AFN:prop-domN-domX}, \ref{AFN:prop-domN-domG}
and the second assertion is a 
well-known property of the square root of a positive operator (cf.\ Thm.\ V.3.24 of Ref.\ \cite{kato}). 

Property (2) follows from the inclusion $\Dom(G^2)\subseteq\Dom(G)=\Dom(N)$ and 
$L_\ell(\Dom(N))\subseteq\Dom(N^{1/2})$. 
\end{proof}

\medskip
We now apply the sufficient condition for conservativity taking as the operator
$C$ (cf. Ref. \cite{ChFa}) the self-adjoint operator
\[
Xu = rp^2 + (pq+qp) +(\omega^2r+ 2\gamma)q^2\ \text{on}\ \Dom(X)=\Dom(N)
\]
with $r>{(2\gamma)}^{-1}$.

\begin{proposition}\label{AFN:prop-inequality-Markov}
Suppose that $V$ is twice differentiable and satisfies the growth condition 
(\ref{AFN:potential-cond}). Then there exists a positive constant $b$ such that
\begin{equation}\label{AFN:eq-inequality-Markov}
2\re\langle Xu,Gu\rangle 
+ \sum_{\ell=1}^2\left\Vert X^{1/2}L_\ell u\right\Vert^2
\le b\left\Vert X^{1/2}u\right\Vert^2
\end{equation}
for all $u\in\Dom(N)$.
\end{proposition}

\begin{proof}
We first check the above inequality for $u\in C_c^\infty(\IR)$. The vector $u$ clearly belongs 
to the domain of the operators $XG,G^*X,L_\ell^*XL_\ell$. Therefore, the left-hand side 
is equal to $\langle u, \Ll(X) u\rangle$. 
{}From \eqref{e:16} we have then
\begin{eqnarray}\label{AFN:eq-estimate-LX}
&&\langle u, \Ll(X) u\rangle  =  
-2 \left\langle u, \left((2\gamma r-1)p^2 + \omega^2 q^2\right)u\right\rangle \\ 
&&+ \left(2rD_{pp}+4D_{pq}+2(\omega^2r+ 2\gamma)D_{qq}\right)\left\Vert u\right\Vert^2 
+ i \left\langle u, [V(q),X]u\right\rangle\,. \nonumber
\end{eqnarray}
Estimating the commutator as follows (cf.\ \eqref{e:16})
\begin{eqnarray*}
\left|i\left\langle u, [V(q),X]u\right\rangle\right| 
& = & \left|-\left\langle u, 
    \left(r(pV^\prime(q)+V^\prime(q)p)+2qV^\prime(q)\right)u\right\rangle\rangle\right| \\
& \le & 2r\Vert pu\Vert \cdot\Vert V^\prime(q)u\Vert 
+  2\Vert qu\Vert \cdot\Vert V^\prime(q)u\Vert  \\
& \le & \left\langle u, (r^2p^2+q^2) u\right\rangle 
+  2\left\langle u,\left|V^\prime(q)\right|^2u\right\rangle \\
& \le & \left\langle u, (r^2p^2+(2g_V^2+1)q^2+2g_V^2) u\right\rangle\,,
\end{eqnarray*}
and putting $c_5= \max\left\{\,r^2,2g_V^2 +1, 2rD_{pp}+4D_{pq}+2(\omega^2r+ 2\gamma)D_{qq}\,\right\}$ 
we find
\begin{eqnarray*}
\left|\langle u, \Ll(X) u\rangle \right|& \le & c_5 \left\langle u, \left(p^2+q^2+2\right) u\right\rangle \\
& = & c_5 \left\langle u, \left(2N+3\right) u\right\rangle \\
& \le & b \left\langle u, X u\right\rangle\,.
\end{eqnarray*}
as in the proof of Prop.~\ref{AFN:prop-domN-domX}.  Hence, we have now proved 
the inequality~(\ref{AFN:eq-inequality-Markov})
for $u\in C_c^\infty(\IR)$. The extension to arbitrary $u\in\Dom(N)$ follows by a 
standard approximation argument. 
\end{proof} 

This result yields
\begin{theorem}\label{AFN:th-unital}
Suppose that the potential $V$ is twice differentiable and satisfies 
the growth condition (\ref{AFN:potential-cond}). Then the minimal 
semigroup associated with the above operators $G,L_1,L_2$ is Markov.
\end{theorem}

\begin{proof}
It suffices to apply Theorem~4.4 from Ref. \cite{ChFa}, choosing the positive, self-adjoint operator $\Phi:=-G_0$ 
introduced in the proof of Prop.~\ref{AFN:prop-domN-domG}. The 
hypothesis {\bf C} holds by Propositions~\ref{AFN:prop-dom-hypoC} and  
\ref{AFN:prop-inequality-Markov}, and the hypothesis  {\bf A} by Prop.~\ref{AFN:prop-G-generates}.
Moreover, we choose the positive, self-adjoint operator $C$ as a sufficiently large multiple of $X$ to satisfy
$$
  \langle \Phi^{\frac12}u\,,\,\Phi^{\frac12}u\rangle \le 
  \langle C^{\frac12}u\,,\,C^{\frac12}u\rangle \quad \forall u\in \Dom(X)\,.
$$
\end{proof}



\section{Stationary state}\label{steady}

In order to establish the existence of a steady state of QFP we now start to verify the conditions of Theorem~VI.1 in Ref. \cite{FaRe01}.

\begin{theorem}\label{AFN:th-exists-invariant state} Suppose that $\gamma>0$ 
and the potential $V$ is twice differentiable and satisfies the growth condition \eqref{AFN:potential-cond}.
Then the quantum Markov semigroup (QMS) with form generator 
(\ref{AFN:eq-form-generator}) has a normal invariant state.
\end{theorem}

\begin{proof}
We shall apply Theorem~IV.1 from Ref. \cite{FaRe01}. Hypothesis {\bf (H)} is satisfied due to Prop.~\ref{AFN:prop-G-generates}. Now choose
\[
X:= rp^2 + (pq+qp) + \left(\omega^2 r + 2\gamma\right)q^2\, , \quad 
Y:= c_6 (p^2+q^2) - c_7\mathds{1}
\]
with $r>{(2 \gamma)}^{-1}$ and $c_6,c_7>0$. 
Indeed $X$ is clearly positive (cf.\ Lemma \ref{AFN:lem-op-rs-positive}). $Y$ is bounded below with finite 
dimensional spectral projections associated with intervals $]-\infty,\Lambda]$, since
it is a translation of a multiple of the number operator. 

In order to check the fundamental inequality 
\begin{equation}\label{e:23}
\pounds(X)[u,u] \le -\langle u,Yu\rangle
\end{equation}
for all $u\in\Dom(N)$ we start from the identity (\ref{AFN:eq-estimate-LX}) in 
the proof of Prop.~\ref{AFN:prop-inequality-Markov} and estimate the commutator as follows
\begin{eqnarray*}
\left|i\left\langle u, [V(q),X]u\right\rangle\right| 
& = & \left|-\left\langle u, 
    \left(r(pV^\prime(q)+V^\prime(q)p)+2qV^\prime(q)\right)u\right\rangle\rangle\right| \\
& \le & 2r\Vert pu\Vert \cdot\Vert V^\prime(q)u\Vert 
+  2\Vert qu\Vert \cdot\Vert V^\prime(q)u\Vert  \\
& \le & \epsilon\left\langle u, (p^2+q^2) u\right\rangle 
+ \frac{r^2+1}{\epsilon} \left\langle u,\left|V^\prime(q)\right|^2u\right\rangle \\
 & \le & \epsilon\left\langle u, (p^2+q^2) u\right\rangle 
+ g_V^2\frac{r^2+1}{\epsilon} \left\langle u,(1+|x|^{2\alpha})u\right\rangle\,,
\end{eqnarray*}
where we used the elementary inequality
\[
(1+|x|^2)^\alpha \le 1 + |x|^{2\alpha}
\]
for $0\le\alpha\le 1$. With the Young inequality we have
\[
|x|^{2\alpha} = \eta^{-1}\cdot(\eta|x|^{2\alpha}) 
\le \frac{(\eta|x|^{2\alpha})^{1/\alpha}}{1/\alpha}
+\frac{(\eta^{-1})^{1/(1-\alpha)}}{1/(1-\alpha)}
= \alpha\eta^{1/\alpha}|x|^2 + \frac{1-\alpha}{\eta^{1/(1-\alpha)}}
\]
for all $\eta>0$. Choosing $\eta=(\epsilon^2/r^2+1)^{\alpha}$ we obtain 
\[
|x|^{2\alpha} \le \frac{\alpha\epsilon^2}{r^2+1}|x|^2 
+ \frac{(1-\alpha)(r^2+1)^{\alpha/(1-\alpha)}}{\epsilon^{2\alpha/(1-\alpha)}}\,.
\]
Therefore we have 
\begin{multline*}
\left|i\left\langle u, [V(q),X]u\right\rangle\right| 
\le   \epsilon\left\langle u, (p^2+q^2) u\right\rangle 
+ \alpha\epsilon g_V^2\langle u,q^2 u\rangle \\
+g_V^2\frac{r^2+1}{\epsilon}\left(1+(1-\alpha)\left(\frac{r^2+1}{\epsilon^2}\right)^{\alpha/(1-\alpha)}\right)
 \left\Vert u\right\Vert^2  \\
 \le  (1+\alpha g_V^2) \epsilon\left\langle u, (p^2+q^2) u\right\rangle 
+g_V^2\left\Vert u\right\Vert^2\left( \frac{(1-\alpha)(r^2+1)^{1/(1-\alpha)}}{\epsilon^{(1+\alpha)/(1-\alpha)}} +\frac{r^2+1}{\epsilon}
 \right)\,.
\end{multline*}
This inequality and (\ref{AFN:eq-estimate-LX}) give
\begin{multline*}
\langle u, \Ll(X) u\rangle \le 
-2 \left\langle u, \left((2\gamma r-1)p^2 + \omega^2 q^2\right)u\right\rangle 
+ (1+\alpha g_V^2) \epsilon\left\langle u, (p^2+q^2) u\right\rangle \\
 + \left( 2rD_{pp}+4D_{pq}+2(\omega^2r+ 2\gamma)D_{qq}\right)\left\Vert u\right\Vert^2\\
+g_V^2\left(\frac{(1-\alpha)(r^2+1)^{1/(1-\alpha)}}{\epsilon^{(1+\alpha)/(1-\alpha)}}+\frac{r^2+1}{\epsilon}\right)
 \left\Vert u\right\Vert^2\,.
\end{multline*}
For all $r>(2 \gamma)^{-1}$, we can take an $\epsilon$ small enough such that 
\[
c_6:=2\min\left\{\,(2\gamma r-1),\omega^2\,\right\}-(1+\alpha g_V^2) \epsilon >0\,.
\]
Putting  
\[
c_7:=2rD_{pp}+4D_{pq}+2(\omega^2r+ 2\gamma)D_{qq}
+g_V^2\frac{(1-\alpha)(r^2+1)^{1/(1-\alpha)}}{\epsilon^{(1+\alpha)/(1-\alpha)}}+g_V^2\frac{r^2+1}{\epsilon}
\] 
we find the asserted inequality \eqref{e:23}:
\[
\langle u, \Ll(X) u\rangle  \le -c_6\langle u, (p^2+q^2)u\rangle 
+ c_7 \left\Vert u\right\Vert^2
\]
for $u\in C_c^\infty(\IR)$. The extension to arbitrary $u\in\Dom(N)$ follows by a 
standard approximation argument. 

Clearly $G$ is relatively bounded with respect to $X$ by previous results {(cf.\ \S4,5).}
Moreover we have $(\lambda+X)^{-1}(\Dom(N))=\Dom(N^2)$, and hence
\[
L_\ell\left((\lambda+X)^{-1}\,\Dom(N)\right)\subseteq L_\ell(\Dom(N))\subseteq\Dom(N^{1/2})=\Dom(X^{1/2})
\]
for $\ell=1,\,2;\;\lambda\ge1$.
Therefore Theorem IV.1\cite{FaRe01} can be applied and it yields the existence of a normal invariant state. 
\end{proof} 
\medskip
We remark that the operators $X$ and $Y$ do not commute here, in contrast to most examples in Ref. \cite{FaRe01}.



\section{Irreducibility and large time behavior}\label{AFN:sect-irreduc}

A QMS $\Tt$ on ${\mathcal{B}}(\initsp)$ is called {\em irreducible} 
if the only subharmonic projections $\Pi$ in $\initsp$ 
({\it{i.e.}} projections satisfying $\Tt_t(\Pi)\ge \Pi$ for all $t\ge 0$) 
are the trivial ones $0$ or $\mathds{1}$. 

If a projection $\Pi$ is subharmonic, the total mass of any normal state 
$\sigma$ with support in $p$ ({\it{i.e.}} such that $\Pi\sigma \Pi = \Pi\sigma =\sigma \Pi
= \sigma$), remains concentrated in $\Pi$ during the evolution. Indeed, 
the state $\Tt_{*t}(\sigma)$ at time $t$ then satisfies
\[
1=\tr{\Tt_{*t}(\sigma)}\ge \tr{\Tt_{*t}(\sigma)\Pi}=\tr{\sigma \Tt_t(\Pi)}
\ge \tr{\sigma \Pi}=\tr{\sigma}=1\, .
\]
As an example, the support projection of a normal stationary state for a QMS is subharmonic (cf.\ Th. II.1 in Ref. \cite{FaRe02}).
Thus if a QMS is irreducible and has a normal invariant 
state, then its support projection must be $\mathds{1}$, {\it{i.e.}} it must be 
faithful. 

In this section we shall prove that the QMS we constructed in \S\ref{QDS}-\ref{s.6} is irreducible 
if the strict inequality $\Delta>0$
holds. The more delicate limiting case $\Delta=0$ is postponed to the next section.

Subharmonic projections are characterized by the following theorem \cite{FaRe02}. 

\begin{theorem}\label{AFN:th-char-subharm-proj}
A projection $\Pi$ is subharmonic for the QMS associated with the operators 
$G,L_\ell$ if and only if its range $\Xx$ is an invariant subspace 
for all the operators $P_t$ of the contraction semigroup generated by $G$ 
({\it{i.e.}} $\forall t\geq 0: P_t{\mathcal{X}}\subseteq{\mathcal{X}}$) and 
\[
L_\ell\left(\Xx\cap\Dom(G)\right)\subseteq {\Xx}
\]
for all $\ell$'s.
\end{theorem}

The application to our model yields the following
\begin{theorem}\label{AFN:th-irreducible}
Suppose that $\Delta>0$.
Then the QMS $\Tt$ associated with 
(the closed extensions of) the operators $G,L_\ell$ given by (\ref{AFN:eq-G}) and (\ref{AFN:Lindblad-Operators}) is irreducible. 
\end{theorem}

\begin{proof}
Let $\Pi$ be a subharmonic projection with range $\Xx\subseteq\initsp$. We shall prove that either $\Pi=0$ or $\Pi=\mathds{1}$, and hence $\Tt$ is irreducible.\\

The domain of $N$ is $P_t$ invariant because it 
coincides with $\Dom(G)$ which is obviously $P_t$-invariant. Since 
both $L_1,L_2$ map $\Dom(N)$ into $\Dom(N^{1/2})$ and $\Dom(N^{1/2})$ 
into $\h$, we have
\begin{eqnarray}
L_\ell\left(\Xx\cap\Dom(N)\right) & \subseteq & \Xx\cap\Dom(N^{1/2})\,,\quad\text{and}\notag\\
L_\ell\left(\Xx\cap\Dom(N^n)\right) & \subseteq & \Xx\cap\Dom(N^{n-1/2})\,. \label{AFN:DomLl}
\end{eqnarray}
Then, by the linear independence of $L_1$ and $L_2$ (due to $\Delta>0$)
\[
\begin{split}
&p\left(\Xx\cap\Dom(N)\right)\subseteq\Xx\cap\Dom(N^{1/2})\,,\quad \text{and}\\
&q\left(\Xx\cap\Dom(N)\right)\subseteq\Xx\cap\Dom(N^{1/2})\,,
\end{split}
\]
and thus, since $N=(p^2+q^2 -1)/2$,
\[
N\left(\Xx\cap\Dom(N)\right)\subseteq\Xx\,.
\]
For all $n>0$ the resolvent operator $R(n;G)=(n-G)^{-1}$ 
maps $\h$ in $\Dom(G)=\Dom(N)$, therefore the operator 
$N_n=n N R(n;G)$ is defined on $\h$. It is also 
bounded, because $N$ is relatively bounded with respect 
to $G$ and the identity $GR(n;G)u=-u +n R(n;G)u$ yields 
the inequalities 
\begin{eqnarray*}
\left\Vert N_n u \right\Vert 
& \le & c\left\Vert n G R(n;G)u\right\Vert 
+ c\left\Vert n R(n;G)u\right\Vert  \\
& \le & cn \left\Vert u\right\Vert 
+ c(n +1)\left\Vert n R(n;G)u\right\Vert,
\end{eqnarray*}
for all $u\in\h$.

The subspace $\Xx$ is clearly $N_n$--invariant because  
\begin{equation*}
R(n;G)=\int_{0}^{\infty}\exp(-nt) P_t\,dt 
\end{equation*}
maps it into $\Xx\cap\,\Dom(G)=\Xx\cap\,\Dom(N)$ 
and $N(\Xx\cap\,\Dom(N))\subseteq\Xx$. It follows that $\Xx$ 
is invariant for the operators $\hbox{\rm e}^{-tN_n} 
=\sum_{k\ge 0}(-tN_n)^k/k!$ ($t\ge 0$) of the (semi)group 
generated by $N_n$.

Notice that, for all $u\in\Dom(N)=\Dom(G)$, we have
\begin{eqnarray*}
\left\Vert N_n u - Nu\right\Vert
& = & \left\Vert N\left( nR(n;G)u-u\right)\right\Vert \\
& \le & c\left\Vert  G \left( nR(n;G)u-u\right)\right\Vert 
+ c\left\Vert  nR(n;G)u-u\right\Vert \\
& = & c\left\Vert  nR(n;G)Gu-Gu\right\Vert 
+ c\left\Vert  nR(n;G)u-u\right\Vert.
\end{eqnarray*}
Thus, by the well-known properties of the Yosida approximations,
$N_n u$ converges towards $Nu$ as $n$ tends to infinity for all 
$u\in\Dom(N)$. It follows then from a well-known result in 
semigroup theory (cf. \cite{kato}, Chapter IX) that the operators $\hbox{\rm e}^{-tN_n}$ 
converge towards the operators $\hbox{\rm e}^{-tN}$ of the 
semigroup generated by $-N$ in the strong operator topology 
on $\h$ uniformly for $t$ in bounded subsets of $[0,+\infty[$. 
Therefore $\Xx$ is $\hbox{\rm e}^{-tN}$--invariant for all $t\ge 0$. 

Since the operators $\hbox{\rm e}^{-t N}, t>0$ are compact 
and self-adjoint, it follows that $\Xx$ is generated by eigenvectors of $N$,
\[
\Xx = \overline{ \hbox{\rm Lin}\left\{\, e_{j}\,\mid\, j\in J\,\right\} }\,,
\]
for some $J\subseteq\mathbb{N}_0$. Here $(e_j)_{j\ge 0}$ is the canonical orthonormal 
basis of $\ell^2(\mathbb{N}_0)$. Moreover, using $\Xx_\infty:=\Xx\cap_{n\ge 1}\Dom(N^n)$, we have $\hbox{\rm Lin}\left\{\, e_{j}\,\mid\, j\in J\,\right\}\subseteq\Xx_{\infty}\subseteq\Xx$, since $e_{j}$ are the eigenvectors of $N$.

By \eqref{AFN:DomLl}, $\Xx_{\infty}$ is $L_1,L_2$--invariant. Since $a$ and $a^\dagger$ are related to 
$L_1,L_2$ by an invertible linear transformation, $\Xx_{\infty}$ is also $a, a^\dagger$--invariant. Therefore, if $J$ is not 
empty ({\it{i.e.}} $\Xx\neq\{\,0\,\}$) taking $m=\min J$ we find immediately
that $e_0=a^{m}e_m/\sqrt{m!}$ belongs to $\Xx_{\infty}\subseteq\Xx$. Thus, any $e_k$ with 
$k>0$ belongs to $\Xx_\infty$ because $a^{\dagger k}e_0=\sqrt{k!}e_k$ and 
$\Xx=\Xx_\infty$ coincides with the whole of $\initsp$. 
\end{proof}

\begin{corollary}\label{AFN:inv-state-faithfull}
If $\Delta>0$, then all normal invariant states are faithful.
\end{corollary}

\begin{proof}
The support projection of an invariant state 
is subharmonic (and non-zero). Since the QMS is irreducible, all 
non-zero subharmonic projections must coincide with the identity 
operator $\mathds{1}$. Hence, any normal invariant state must be faithful. 
\end{proof}

\medskip
Recall the following classical result \cite{Fr}.
\begin{theorem}
Let $\Tt$ be the unital minimal QMS associated with operators $G, L_\ell$.
Suppose that $\Tt$ has a faithful normal invariant state $\rho$. Then the 
vector space of fixed points 
\[
\Ff(\Tt)=\set{x\in{\mathcal{B}(\initsp})\,\mid\, \Tt_t(x)=x,\,\forall t\ge 0}
\]
is an algebra and the invariant state $\rho$ is unique if and only if 
$\Ff(\Tt)={\mathbb{C}}\mathds{1}$. 
\end{theorem}

Remember that the QMS under consideration admits a steady state by Theorem~\ref{AFN:th-exists-invariant state} and by Corollary~\ref{AFN:inv-state-faithfull} this invariant state is 
faithful provided that we have strict inequalitiy in the Lindblad 
condition \eqref{AFN:Lindblad-cond}.

The next result, taken from Proposition~2.3 in Ref. \cite{FaRe98} and the Correction in Ref. \cite{FaRe08}, allows us to apply immediately the above 
theorem and show that the QMS converges towards its unique invariant 
state. Let us introduce first
\[
\Nn(\Tt)=\set{x\in{\mathcal{B}(\initsp})\mid 
\Tt_t(x^*x)=\Tt_t(x^*)\Tt_t(x)\,,
\Tt_t(xx^*)=\Tt_t(x)\Tt_t(x^*) 
\,\forall t\ge 0}\,.
\]
If $\Tt$ has a {\em faithful} normal invariant state, it is easy to show 
by the Schwarz property $\Tt_t(x^*)\Tt_t(x)\le \Tt_t(x^*x)$, that $\Nn(\Tt)$ 
is an algebra.

\begin{theorem}\label{AFN:thm-convergene}\cite{FaRe98,FaRe08}
Let $\Tt$ be the unital minimal QMS associated with the operators $G, L_\ell$.
Suppose that: i) there exists a domain $D_c$ which is a core for both 
$G$ and $G^*$, and ii) for all $u\in D_c$, its image $R(n;G)u$ belongs to $\Dom(G^*)$ 
and the sequence $(nG^*R(n;G)u)_{n\ge 1}$ converges strongly.

Then
\begin{enumerate}
\item $\Ff(\Tt)=\set{H,L_1,L_1^{*},L_2,L_2^{*}}^\prime$\,,
\item $\Nn(\Tt)\,\subseteq\,\set{L_1,L_1^{*},L_2,L_2^{*}}^\prime$\,,
\item if $\Ff(\Tt)=\Nn(\Tt)$ then for all initial states $\sigma$, $\Tt_{*t}(\sigma)$ 
converges as $t$ goes to infinity towards an invariant state in the trace norm.
\end{enumerate} 
\end{theorem}

In the above Theorem the set $\{H,L_1,L_1^*,L_2,L_2^*\}'$ denotes the {\em commutant}, {\it i.e.} the set of all 
operators that commute with $H$ as well as with $L_1$, $L_1^*$ and $L_2$, $L_2^*$ 
(analogously for $\{L_1,L_1^{*},L_2,L_2^{*}\}'$) and $R$ denotes the resolvent. 
Because $L_1$ and $L_2$ are linearly independent, as long as 
we have strict inequality in \eqref{AFN:Lindblad-cond}, we see 
that $\{L_1,L_2\}'$ consists only of operators commuting with 
$q$ and $p$. This yields $\Nn(\Tt)=\mathbb{C}\mathds{1}$ and since $\mathbb{C}\mathds{1}\subseteq\Ff(\Tt)\subseteq\Nn(\Tt)$ also $\mathbb{C}\mathds{1}=\Ff(\Tt)=\Nn(\Tt)$. 

To apply Theorem~\ref{AFN:thm-convergene} and thus 
ensure convergence we need to verify the conditions 
$(i)$ and $(ii)$. Condition $(i)$ is obvious 
from Prop.~\ref{AFN:prop-domN-domG} (one might pick 
$C_c^\infty(\mathbb{R})$ as a core). 
Condition $(ii)$ is also easily checked 
because the operators $G$, $G^*$ and $N$ have the same 
domain by Prop.~\ref{AFN:prop-domN-domG}. Moreover, their 
graph norms are equivalent (cf. Remark~\ref{AFN:graph}). 
Now, for all $u\in D_c=C_c^\infty(\mathbb{R})$, since  
$nGR(n;G)u = nR(n;G)Gu$, the sequence $(nR(n;G)u)_{n\ge 1}$ 
converges to $u$ in the graph norm of $G$. By Remark~\ref{AFN:graph} it is also convergent in the graph norm of $G^*$, {\it{i.e.}} 
$(nG^*R(n;G)u)_{n\ge 1}$ converges to $G^*u$.

Altogether we have proved the following
\begin{corollary}\label{AFN:cor-conv-SS}
Let $\gamma>0$ and $V\in C^2(\mathbb{R})$ satisfy \eqref{AFN:potential-cond}. If $\Delta>0$ the QMS associated with $G$ 
and $L_\ell$ has a unique faithful normal invariant state $\rho$. 
Moroever, for all normal initial states $\sigma$, we have
\[
\lim_{t\to\infty}\Tt_{*t}(\sigma) = \rho
\]
in the trace norm.
\end{corollary}



\section{The limiting case $\Delta=0$}
\label{AFN:sect-limiting}

Here we study the case when $\Delta=D_{pp}D_{qq}-D_{pq}^2-\gamma^2/4=0$ and $\gamma>0$.
We start with the case $V=0$. In this situation we can compare our result to the explicit formula for the (unique normalized) steady state in Ref. \cite{SCDM} for zero perturbing potential.
The kernel of the density matrix of the steady state can be calculated by means of Fourier transform\cite{SCDM} and is given by
\begin{equation}\label{AFN:steadystate-expl}
\begin{split}
\rho_\infty(x,y)=\frac{\gamma\omega}{\pi\sqrt{\gamma Q_{22}}} & \exp\left(-\frac{1}{4\gamma Q_{22}}\left[\gamma^2\omega^2(x+y)^2+Q(x-y)^2\right]\right)\\
\times & \exp\left(-i\omega\frac{Q_{12}}{Q_{22}}\left(\frac{x^2-y^2}{2}\right)\right)\,,
\end{split}
\end{equation}
where the abbreviations are 
\begin{eqnarray*}
Q_{11}&=& D_{pp}+\omega^2 D_{qq}\\
Q_{12}&=& 2\omega\gamma D_{qq}\\
Q_{22}&=& D_{pp}+\omega^2 D_{qq}+4\gamma(D_{pq}+\gamma D_{qq})\quad \text{and}\\
Q&=& Q_{11}Q_{22}-Q_{12}^2\,.
\end{eqnarray*}
One can see that this becomes a \emph{pure} state if and only if $Q=\gamma^2\omega^2$. We will now discuss its implications on the relation between diffusion and friction coefficients.

\begin{lemma}\label{AFN:lem-V=0} 
Let $V=0$, $\gamma>0$, and the Lindblad condition hold.
The steady state given by \eqref{AFN:steadystate-expl} is a pure state, {\it{i.e.}} $Q=\gamma^2\omega^2$, if and only if $\gamma<\omega$ and
\begin{eqnarray}
0 & = & D_{pp}D_{qq}-D_{pq}^2-\gamma^2/4 \label{AFN:lindblad0}\\
D_{pq} & = & -\gamma D_{qq} \label{AFN:limitcase-Dpg}\\
D_{pp} & = & \omega^2 D_{qq}\,.\label{AFN:limitcase-Dpp}
\end{eqnarray}
In this case 
\begin{eqnarray}
D_{qq} & = & \frac{\gamma}{2\sqrt{\omega^2-\gamma^2}}\, ,\quad \text{and}\notag\\
\rho_\infty(x,y) & = & \frac{1}{\pi}\sqrt[4]{\omega^2-\gamma^2}\mbox{e}^{-(cx^2+\overline{c}y^2)}\,\quad \text{with}\label{AFN:steadysimple}\\
c & = & \frac{1}{2}(\sqrt{\omega^2-\gamma^2}+i\gamma)\,.\label{AFN:constant}
\end{eqnarray}
\end{lemma}
\begin{proof}
We rewrite
\begin{eqnarray}
0 & = & Q-\gamma^2\omega^2=\left(1-\frac{\gamma^2}{\omega^2}\right)\left(D_{pp}-D_{qq}\omega^2\right)^2\label{AFN:pure1}\\
& + & \frac{\gamma^2}{\omega^2}\left(D_{pp}+2D_{pq}\frac{\omega^2}{\gamma}+D_{qq}\omega^2\right)^2\label{AFN:pure2}\\
& + & 4\omega^2\left(D_{pp}D_{qq}-D_{pq}^2-\frac{\gamma^2}{4}\right)\label{AFN:pure3}\, .
\end{eqnarray}
In the case $\gamma<\omega$ this directly implies \eqref{AFN:lindblad0}--\eqref{AFN:limitcase-Dpp}, because of \eqref{AFN:Lindblad-cond}.\\
For $\gamma=\omega$ the term \eqref{AFN:pure2} is a quadratic polynomial in $\omega$. But it has no real zero, since $D_{pp}D_{qq}-D_{pq}^2>0$ by \eqref{AFN:pure3}.\\
For the case $\gamma>\omega$ we rewrite \eqref{AFN:pure1}--\eqref{AFN:pure3} as
\begin{eqnarray}
0 & = & \left(D_{pp}+2D_{pq}\gamma+D_{qq}\omega^2\right)^2 \label{AFN:pure21}\\
& + & 4\gamma^2\left(D_{pp}D_{qq}-D_{pq}^2-\frac{\gamma^2}{4}\right)\label{AFN:pure22}\\
& + & \gamma^2(\gamma^2-\omega^2)\,,\label{AFN:pure23}
\end{eqnarray}
which is a contradiction since the summand \eqref{AFN:pure22} is non--negative by \eqref{AFN:Lindblad-cond}.\\
The form of $D_{qq}$ and the kernel of the density matrix of the steady state follow by straightforward calculations.
\end{proof}
\begin{corollary} Let $V=0$, $\Delta=0$, and $0<\gamma<\omega$. Under the conditions \eqref{AFN:lindblad0}, \eqref{AFN:limitcase-Dpg} and \eqref{AFN:limitcase-Dpp} of Lemma \ref{AFN:lem-V=0} the semigroup is not irreducible. It admits a steady state that is not faithful.
\end{corollary}
We now want to interpret this result in terms of the generators.
Condition \eqref{AFN:lindblad0} implies that the Lindblad operators $L_1$ and $L_2$ are no longer linearly independent since $L_2$ is zero. The operator $L_1$ has a nontrivial kernel containing functions of the type $\exp(-c x^2)$, with $c$ from \eqref{AFN:constant}, which are clearly also in the kernel of $L_1^*L_1$. Hence, density matrices with kernel \eqref{AFN:steadysimple} are initially not affected by the dissipative part of the evolution. However the Hamiltonian evolution might move them away from the kernel. It is exactly conditions \eqref{AFN:limitcase-Dpg} and \eqref{AFN:limitcase-Dpp} that ensure that $H$ is (up to an additive constant) a scalar multiple of $L_1^*L_1$ and thus $[H,L_1^*L_1]=0$. Of course in this case functions of type \eqref{AFN:steadysimple} stay unaffected by the evolution.\\

Using the explicit formula (\ref{AFN:steadystate-expl}) for the kernel of the density matrix in the case $V=0$ one can check the non--faithfulness of the steady state also directly:
\begin{remark}
For $\rho$ \emph{not}\, to be faithful we have to find a $u\in \initsp$ with $u\neq0$, such that $\left<u,\rho u\right>=0$. By positivity of $\rho$  this is equivalent to $\rho u=0$. Using \eqref{AFN:steadystate-expl}, such a $u$ satisfies 
\begin{equation*}
\begin{split}
 0=\int_{\mathbb{R}}&\exp\left(-\frac{1}{4\gamma Q_{22}}\left[\gamma^2\omega^2(x+y)^2+Q(x-y)^2\right]\right)\\
\times & \exp\left(-i\omega\frac{Q_{12}}{Q_{22}}\left(\frac{x^2-y^2}{2}\right)\right) u(y)dy\, ,
\end{split}
\end{equation*}
for all $x\in\mathbb{R}$. Looking pointwise in $x$ we can drop the factors 
depending only on $x$ and get
\begin{equation*}
\begin{split}
0=\int_{\mathbb{R}} & \exp \left(-\frac{\gamma^2\omega^2-Q}{2\gamma Q_{22}}xy\right)\\
& \times\exp\left(-\frac{1}{4\gamma Q_{22}}\left(\gamma^2\omega^2+Q-2i\omega\gamma Q_{12}\right)y^2\right) u(y)dy\,.
\end{split}
\end{equation*}
Provided the steady state is not pure, {\it i.e.} $Q\neq\gamma^2\omega^2$, one can interpret this as the Laplace--transform of the term in the second line. Due to the decay of the exponential factor it exists for all $x\in\mathbb{R}$ and it can not be zero for all $x$ unless $u$ is zero. This shows that the steady-state $\rho$ is faithful except when $Q=\gamma^2\omega^2$. In the case of a pure steady state the first exponential becomes one and the above integral vanishes for all functions that are orthogonal to 
\begin{equation*}\label{AFN:kernel-L1}
\exp\left(-\frac{1}{4\gamma Q_{22}}\left(\gamma^2\omega^2+Q-2i\omega\gamma Q_{12}\right)y^2\right)=\exp(-\overline{c}y^2)\,.
\end{equation*}
Note that the complex conjugate of this function is in the kernel of $L_{1}$, as noted earlier. 
\end{remark}

The above discussion on the explicit steady 
state for $V=0$ and algebraic computations of the invariant subspaces 
for $G$ and $L_1$ showing that they should be trivial, lead us to the 
following

\noindent{\bf Conjecture.} When $\Delta=0$ but $V\not=0$ (with $V\in C^2(\mathbb{R})$ satisfying \eqref{AFN:potential-cond}) the semigroup 
has a unique faithful normal invariant state and the conclusion of 
Corollary \ref{AFN:cor-conv-SS} holds.

The full proof of this conjecture, however, entails a lot of analytical 
difficulties and technicalities on invariant subspaces for strongly 
continuous semigroups and their perturbations and will be postponed 
to a forthcoming paper.


\bigskip

\noindent
{\bf Acknowledgement:}
A.~A. acknowledges partial support from the DFG under Grant No. AR
277/3--3, the ESF in the project ``Global and geometrical aspects of
nonlinear partial differential equations'', and the
Wissenschafts\-kolleg \emph{Differentialgleichungen} of the FWF. \\
F.~F. acknowledges partial support from the 
GNAMPA group project 2008 ``Quantum Markov Semigroups''. He also 
acknowledges the Institute for Analysis and Scientific Computing, 
TU Wien, were a part of this paper was written, for warm hospitality.\\
The authors are grateful to the organizers of the 
$28^{th}$ conference on ``Quantum probability and related topics'', Guanajuato, Mexico, and in particular to Roberto Quezada.


\bibliographystyle{amsplain}  

\end{document}